# Strong mid-infrared photoresponse in small-twist-angle bilayer graphene

Bingchen Deng[1†], Chao Ma[1†], Qiyue Wang[2†], Shaofan Yuan[1†], Kenji Watanabe[3], Takashi Taniguchi[3], Fan Zhang[2]* and Fengnian Xia[1]*

[1]*Department of Electrical Engineering, Yale University, New Haven, Connecticut 06511, USA*

[2]*Department of Physics, The University of Texas at Dallas, Richardson, TX 75080, USA*

[3]*National Institute for Materials Science, 1-1 Namiki, Tsukuba 305-0044, Japan*

[†]These authors contribute equally to this work.

*Correspondence to: fengnian.xia@yale.edu; zhang@utdallas.edu.

**ABSTRACT**

Recently the small-twist-angle (< 2°) bilayer graphene has received extraordinary attentions due to its exciting physical properties[1, 2, 3, 4, 5, 6, 7, 8, 9, 10, 11]. Compared with monolayer graphene, the Brillouin zone folding in twisted bilayer graphene (TBG) leads to the formation of superlattice bandgap and significant modification of density of states[4, 6, 7, 12, 13]. However, these emerging properties have rarely been leveraged for the realization of new optoelectronic devices. Here we demonstrate the strong, gate-tunable photoresponse in mid-infrared wavelength range of 5 to 12 μm. A maximum extrinsic photoresponsivity of 26 mA W$^{-1}$ has been achieved at 12 μm when the Fermi level in a 1.81° TBG is tuned to its superlattice bandgap. Moreover, the strong photoresponse critically depends on the formation of superlattice bandgap, and it vanishes in the gapless case with ultrasmall twist angle (< 0.5°). Our demonstration reveals the promising optical properties of TBG and provides an alternative material platform for tunable mid-infrared optoelectronics.

**INTRODUCTION**

Recently, various twisted bilayer structures have been extensively explored because they host many novel physical phenomena such as superconductivity and emerging topological properties[6, 7, 8, 9, 10, 11, 12, 13, 14, 15, 16, 17, 18, 19]. Among these twisted bilayer systems, small-twist-angle bilayer graphene represents a particularly interesting material system. First, the emergence of moiré pattern in twisted bilayer graphene (TBG) leads to the formation of mini-Brillouin zone and significantly enhanced dynamical conductivity in the low energy range[20, 21, 22]. Second, the periodic modulation of the interlayer potential in moiré superlattice can induce bandgaps in both electron and hole branches[4, 12, 13]. Third, because the two constituent graphene layers have exactly the same crystalline structure, TBG exhibits interesting physical properties even when the twist angle is large[23, 24]. This phenomenon is very different from graphene/hexagonal boron nitride (hBN) heterostructures, in which the heterostructure properties are strongly modified from those of the constituent layers when the crystalline directions of both constituent layers are aligned[25, 26, 27]. However, previous light-matter interaction experiments in TBG focused on the cases with relatively large twist angles using the visible or near-infrared photons[24, 28, 29, 30, 31, 32], and in these demonstrations van Hove singularity-like electronic resonances were leveraged for enhanced light-matter interactions[23, 28, 29, 31, 32]. In this work, we report strong broadband mid-infrared photoresponse in small-twist-angle TBG. Furthermore, we clarify its bolometric origin through transport studies and reveal the significance of the superlattice bandgap in photoresponse.

**Main Text**

Fig. 1a is the schematic of the TBG transistor used for mid-infrared photodetection in this work. The active channel consists of a hBN/TBG/hBN heterostructure, which is assembled using a dry transfer approach reported in ref. 33. A silicon back gate is used to control the Fermi level in TBG.

We present the information on the detailed fabrication process in Methods. Before we performed the photocurrent measurements, we first characterized the transport properties of the TBG transistor. Fig. 1b plots a transistor source-to-drain resistance as a function of the back-gate bias ($V_{BG}$) measured at 83 K with a source-drain bias ($V_{DS}$) of 10 mV. The thicknesses of the top and bottom hBN layers are 30 and 25 nm respectively in this device and the optical micrograph of this device is shown in the inset of Fig. 1c. Three resistance maxima are observed at $V_{BG}$ of ±43.5 V and around 0 V, respectively. The resistance maximum at $V_{BG}$ of around 0 V (charge-neutrality point) is due to the presence of Dirac point in TBG[4, 5, 34]. The gate voltage at which charge-neutrality occurs ($V_{CNP}$) is always within ±2 V $V_{BG}$ for all the devices in this work and is subtracted in the bottom x-axis in Figs. 1b and 1c. The resistance peaks at ±43.5 V are due to the formation of the superlattice bandgap above and below the lowest moiré Dirac bands, as illustrated in the calculated bandstructure in the inset of Fig. 1b[4, 34]. When the Fermi level is tuned to the centre of the superlattice bandgap, the carrier density in the channel is minimized, leading to the observed resistance peaks. As reported previously[4, 34], 4 electrons per moiré unit cell ($4n_0$) are required to fill/vacate the lowest moiré Dirac bands. We can therefore index the electron filling number in the top x-axis in Fig. 1b and infer the unit cell area (thus the twist angle) of the device from the transport properties. Taking into account the thicknesses of the gate dielectric layers (25 nm hBN and 90 nm SiO$_2$), we calculate that the superlattice gaps are located at $n_{s\pm} = \pm 7.62 \times 10^{12}$ cm$^{-2}$ and the superlattice unit cell area $A = 4/n_{s+} = 52.5$ nm$^2$. Since $A = \frac{\sqrt{3}}{2}\left(\frac{a}{2\sin\frac{\theta}{2}}\right)^2$ and $a = 0.246$ nm (the graphene lattice constant), a twist angle $\theta$ of 1.81° is obtained.

After the characterization of the gate dependent resistance of the TBG device, we measured the gate dependent photocurrent at 5.0, 7.7 and 12 μm, respectively. In the photocurrent measurements, we kept the source-drain bias at 200 mV and swept the $V_{BG}$. The photocurrent was determined by comparing the source-drain current ($I_{DS}$) with and without light illumination based on a lock-in scheme. The detailed information on the photocurrent measurements is included in Methods. As shown in Fig. 1c, the photocurrent results show almost identical gate dependent trends at all three wavelengths. Moreover, there are a few salient features in the photoresponse. First, a strong photoresponse is observed when the device Fermi level is within the superlattice bandgap. At 12 μm, the extrinsic responsivity ($R_{ex}$) reaches 26 mA W$^{-1}$, which is strikingly large, since the channel only consists of two layers of carbon atoms. In fact, such an extrinsic responsivity is comparable to some of the early quantum well infrared photodetectors (QWIP) leveraging the intersubband transitions in the mid-infrared[35, 36]. However, those QWIPs usually consist of tens of pairs of quantum well structures. Second, the photocurrent exhibits both polarities. This observation indicates that the channel can be both more and less conductive under the light illumination, depending on the position of the gate-controlled channel Fermi level. When the Fermi level is tuned to the centre of a superlattice bandgap, a strong, positive photocurrent is observed. However, the photocurrent turns to negative quickly when the Fermi level deviates from the superlattice bandgap. Third, the shapes of resistance in Fig. 1b and $R_{ex}$ in Fig. 1c are very similar, except that the photocurrent has both polarities while the source-drain resistance cannot be negative. In addition, there is electron-hole asymmetry in both the resistance and photocurrent measurements (Fig. 1b and 1c), due to the intrinsic asymmetry in bandstructures. Although there are superlattice-induced bandgaps in both the electron and hole branches, the magnitudes of the bandgaps and the detailed bandstructures are different. Several previous experimental works suggest that the bandgap in the electron side (~60 meV) is slightly larger than that in the hole side (~50 meV) for TBG with a twist angle of around 1.8° (refs. 7, 34, 37). As we will elucidate later, the photoresponse depends on the magnitude of the bandgaps and the detailed bandstructures. As a result, it is not surprising to observe the asymmetry in transport and photoresponse characteristics for all three wavelengths.

In order to clarify the origin of the photoresponse and its gate dependence, we performed the absorption calculations in TBG. In Fig. 2a, we plot the calculated real part of TBG dynamical conductivity $\sigma(\omega)$ (in the unit of monolayer graphene conductivity $\sigma_0$) with a twist angle of 1.81° when the Fermi level of the TBG is tuned to the middle of the superlattice bandgap on the electron side. In the calculation, we only consider the conductivity due to direct interband transitions. We present the details of the calculation in the Methods. In the inset of Fig. 2a, we plot the possible interband transition pathways for photons at 5.0 (purple), 7.7 (blue), and 12 μm (red), respectively. In these transitions, the mid-infrared photons excite the electrons from the filled moiré bands to the empty moiré bands above. As shown in Fig. 2a, the dynamical conductivity right beyond the bandgap energy is about $8\sigma_0$, indicating a very strong light absorption well exceeds 4.6%, which is the absorption of light in two layers of graphene through interband transitions[38]. Although the conductivity becomes smaller at high energy range, it is still larger than $2\sigma_0$ in a broad energy range from 25 to 80 meV. The enhancement of the conductivity is due to the enhanced density of states (DOS) close to the top of the moiré Dirac bands and the bottom of the empty bands. In addition, there are also multiple energy ranges in which the conductivity is beyond $2\sigma_0$. This feature is a consequence of the presence of optical transitions between multiple pairs of moiré bands. Here we want to stress that the calculated dynamical conductivity should not be used to directly compare with the relative photoresponse at different excitation wavelengths, due to the well-known discrepancy between the experimentally measured and theoretically calculated superlattice bandgap values[4, 6, 7, 34, 37, 39]. In general, the calculated superlattice bandgap is only less than half of the experimentally value. Taking this factor into account, most likely at 12 μm (~100 meV), the conductivity is greater than $2\sigma_0$, which explains the strong photoresponse at this wavelength.

A strong absorption of light does not directly transform to the strong photocurrent. In our devices, the photocurrent can arise from either the intrinsic photoconductive effect or the bolometric effect. The former is due to the extra photocarriers induced by light. These additional photocarriers make the material more conductive, and hence the photocurrent is positive. For the bolometric effect, the photocurrent is due to the change of the temperature induced by light. Bolometric effect can lead to both positive and negative photocurrents, depending on the material properties. In metals and superconductors, bolometric effect usually results in a negative photocurrent, since enhancement of temperature leads to larger resistance[40, 41]. In intrinsic semiconductors, higher temperature can lead to reduced resistance due to thermal excitation of carriers[41]. Here we performed the temperature-dependent transport characterizations and the results are plotted in Fig. 2b. The measurements were performed on a different 1.81° TBG device of the same size (3 μm by 3 μm) made from the same hBN/TBG/hBN stack (~20 μm by 20 μm) as the one presented above. When the Fermi level is close to the electron (hole) superlattice bandgap, which corresponds to a $V_{BG}$ of 43.5 (-43.5) V, the channel resistance exhibits clear semiconducting behaviour, and the resistance reduces at higher temperature. On the contrary, at all other gate biases the TBG resistance increases as the temperature rises, which is a typical metallic property. In the inset of Fig. 2b, we plot the temperature coefficient of the conductance (TCC = $\frac{1}{G} \cdot \frac{\Delta G}{\Delta T}$, where $G$ is the device conductance, $T$ is temperature) measured at 83 K. The gate dependence of the TCC resembles that of photoresponse in Fig. 1c clearly, suggesting the bolometric nature of the observed photoresponse in TBG.

In order to further investigate the origin of the photocurrent generation and the reproducibility, we also measured the photocurrent as a function of $V_{DS}$. As shown in Fig. S1c in Supplementary Section I, as the $V_{DS}$ increases, the photocurrent increases initially and then starts to decrease at $V_{DS}$ of merely around 0.1 V. This is the typical behaviour of a bolometric device, as the current heats up the device and reduces the bolometric photoresponse[42]. On the contrary, for a photoconductive detector, the photocurrent can saturate as the bias increases due to the carrier velocity saturation. In our device, the photoresponse starts to decrease at a biasing field of around 0.03 V µm$^{-1}$ (0.1 V source-drain bias in 3 µm-long device) as shown in Fig. S1c. Such a decrease cannot be due to the carrier velocity saturation, since in the transport measurements as shown in Fig. S1d, the drain current does not decrease with biasing field up to 0.17 V µm$^{-1}$ (corresponding to $V_{DS}$ of 0.5 V in this 3-µm long device). As a result, this observation of photoresponse reduction at a small source-drain biasing field further confirms the bolometric effect origin. Besides, in Supplementary Section II, we present frequency- and power-dependent photocurrent measurements in our TBG devices. The photocurrent does not degrade up to the modulation frequency of 10 kilohertz. Moreover, the photocurrent depends linearly on the incident power. These results indicate that photogating effect observed previously in other nanomaterials[43, 44] is not likely to play an important role here.

In addition to the 1.81° TBG, we also measured the photoresponse in a 1.15° TBG. The photoresponse and TCC are presented in the Supplementary Section III. We observed very similar photoresponse properties as those reported in Fig. 1c, as shown in Fig. S3a. In fact, theoretical calculations predict the existence of the superlattice bandgap from 2° to ~1.05° (refs. 4, 12, 13). As a result, it is not surprising that the photoresponse properties are similar in the 1.81° and 1.15° TBG devices. In the 1.15° TBG device, similar behaviour of photoresponse decrease at slightly high $V_{DS}$ is also observed (see Fig. S3b and S3c). Besides, the peak responsivities of the 1.81° TBG device are in general higher than those of the 1.15° case. Two factors could contribute to this observation. First, the superlattice-induced bandgap (at 4 electron/hole per moiré) for 1.81° TBG is larger than that in 1.15° TBG[7], but they are still smaller than the lowest photon energy in our experiment, which is around 100 meV (12 μm light). As shown in Fig. 2a, for photons with energy closer to the bandgap, the optical absorption is higher. Second, the conductance for the 1.81° TBG (Fig. 2b) when the Fermi level is inside the superlattice bandgap at 83 K is higher than that of the 1.15° TBG (Fig. S4a). For a bolometer, since the photocurrent can be expressed as $I_{\mathrm{ph}} = V_{\mathrm{DS}} \cdot G \cdot \mathrm{TCC} \cdot \Delta T$ (ref. 45, where $\Delta T$ is the effective carrier temperature change), the higher conductance in the 1.81° TBG can also lead to larger photoresponse.

Here we also want to emphasize that, in the photocurrent estimation above, we do not distinguish the difference between the electron and phonon temperatures. In our devices the electron and phonon temperatures can be different, due to a sizeable $V_{DS}$. TCC presented in the inset of Fig. 2b was measured at a condition close to equilibrium ($V_{DS}$ = 10 mV) where electrons and phonons have almost the same temperature. Nevertheless, the formula $I_{\text{ph}} = V_{\text{DS}} \cdot G \cdot \text{TCC} \cdot \Delta T$ still captures the physics of photocurrent generation. The positive TCC when the Fermi-level is tuned within the superlattice gap is due to the carrier density dependence on electron temperature. In this case the TBG is close to intrinsic, and the enhanced electron temperature significantly increases the free carrier density, leading to a positive TCC. Away from the superlattice gap, the TBG is doped and has significant amount of free charges. Higher electron temperature does not change the free carrier density significantly. In this case the scattering of free carrier dominates the TCC, which is negative due to the enhanced scattering at higher temperature.

Other than TBG with superlattice bandgap (with twist angle from around 1° to 2°), it is also interesting to investigate the photoresponse in TBG with ultrasmall twist angle, in which previous theoretical and experimental transport characteristics do not indicate a superlattice bandgap opening[4, 9]. Often, the superlattice bandgaps at electron filling states of ±4 close when the twist angle is slightly below 1° and hardly reopen as the twist angle further decreases[11]. We measured the gate-dependent photoresponse in TBG with an ultrasmall twist angle of ~ 0.37°. As plotted in Fig. 3a, the source-drain resistance of such a small twist angle device does not exhibit major side peaks in both electron and hole branches even at 2 K, which is distinctively different from the transport properties reported in TBG with the twist angle of 1.81° (Fig. 1b) and 1.15° (Fig. S4a). We estimate the twist angle in this case by noticing two small side ripples as denoted by the arrows. The two side ripples are positioned at ±1.80 V back-gate voltages (carrier densities of $\pm 3.17 \times 10^{11}$ cm$^{-2}$) relative to the charge-neutrality point, corresponding to electron filling states of ±4 (ref. 9), from which a twist angle of 0.37° is deduced as before. We emphasize that, at these two ripple positions, there is no insulating behaviour and superlattice bandgaps do not exist within the plotted filling range from -32 $n_0$ to 32 $n_0$ (Supplementary Section IV). Since our devices have a two-terminal configuration, contact resistance can smear the intrinsic transport features of the TBG channel. As a result, these ripples are not as evident as those reported previously[9] but are still visible. At such a small twist angle, there is no superlattice bandgap from -32 $n_0$ to 32 $n_0$, as also evidenced by the calculated bandstructure in the inset of Fig. 3a. Note that this nature would not be changed by the complex lattice relaxation and reconstruction[9]. The gate dependent photoresponse of this TBG is plotted in Fig. 3b. No photoresponse peak exists in the electron or hole branches. Moreover, at the Dirac points (the charge neutrality point), we observe weak positive photoresponsivity of around 1.2 mA W$^{-1}$ in this TBG with ultrasmall twist angle. This observation is different from those in TBG with relatively large twist angle where the responsivity does not reach a positive value at the Dirac points. This is probably due to the different bandstructures around the Dirac points when the twist angles are different. In fact, the measured TCC (see inset of Fig. 3b) on this device also supports our measured positive photoresponse at the Dirac points. A small, positive TCC of 0.05% is observed.

In addition, as a comparison, we fabricated AB-stacked bilayer graphene devices sandwiched in hBN layers and measured their mid-infrared photoresponse. We present the results in Supplementary Section V. In short, due to the absence of a superlattice bandgap and the smaller density of states compared with that of TBG, the photoresponse is rather weak (< 1.5 mA W$^{-1}$). It is also worth mentioning that in a previous work by Yan *et al*.[46], dual-gate AB bilayer graphene bolometers have been demonstrated. Their devices operate at 5 K, and the voltage responsivity is as high as ~2×10$^5$ V W$^{-1}$. Given the sample resistance of around 160 kΩ, the voltage responsivity can be converted to current responsivity of around 1250 mA W$^{-1}$. Such a responsivity is much greater than what is reported in our work. However, our devices operate at a higher temperature of around 80 K, at which the dual-gate AB bilayer graphene devices demonstrated by Yan *et al.* are no longer operational. In terms of physical mechanism, bolometric effect is the dominant source of photoresponse in both cases. However, our devices preserve the intrinsic properties of TBG by hBN encapsulation. Moreover, the DOS at the edges of superlattice bandgaps are enhanced due to the folding of Brillouin zone, leading to increased absorption of mid-infrared light especially for the 12 µm light. As a result, the hBN encapsulated TBG represents a new platform for mid-infrared photonics, in addition to AB bilayer graphene.

Finally, we want to comment on the optimum twist angle for mid-infrared photodetection. For light with photon energy below 100 meV or so, the optimum twist angle is likely the one such that the superlattice-induced gap approximately matches the photon energy, due to the strong absorption at the band edge as suggested by the calculations in Fig. 2a. Despite the discrepancy between the calculated and measured superlattice bandgaps, it is expected to be less than ~100 meV (refs. 4, 6, 7, 34, 37, 39). As a result, for photon energy larger than 100 meV, there is no twist angle at which the bandgap matches the photon energy. In this case, the optimum choice is probably the twist angle that leads to the largest gaps. With a relatively large bandgap, the absorption can still be probably enhanced for photon energy below ~200 meV. Moreover, the large bandgap can also lead to enhanced TCC when Fermi level is at the middle of the superlattice bandgap. As reported previously, the superlattice-induced bandgap first increases and then decreases from ~1° to 2° (refs. 4, 34, 37, 39). Beyond 2°, tuning the Fermi-level to the middle of the bandgap is no longer feasible using regular dielectrics as the required doping is overwhelmingly large. According to the previous experimental results[34, 37], the largest superlattice bandgap probably occurs when the twist angle is between 1.4° and 2°. In our experiments, we use three different lasers (12, 7.7 and 5.0 µm), and the photon energy is between 100 and 250 meV. As a result, the device performance should be the best if the superlattice bandgap is the largest. In fact, we fabricated and measured around 10 devices with various twist angles, finding that 1.81° TBG device exhibits the best result (Supplementary Section VI). This observation is consistent with previous experimental works on the superlattice bandgaps of TBGs with different twist angles.

In summary, we report the strong mid-infrared photoresponse in TBG, due to the superlattice-induced bandgap and superlattice-enhanced DOS. Such TBGs show strong photoresponse in a broad mid-infrared wavelength range from 5 to 12 μm, reaching an extrinsic peak responsivity of 26 mA W$^{-1}$ at 12 μm. Moreover, we reveal that the twist angle plays a critical role and such strong photoresponse vanishes when twist angle is ultrasmall, due to the closing of the superlattice-induced bandgap. Our results demonstrate the promising role of TBG in tunable mid-infrared optoelectronic applications.

**Methods**

**Fabrication of twisted bilayer graphene transistors**.

The fabrication of the device started with the mechanical exfoliation of monolayer graphene and hBN flakes on SiO$_2$ covered silicon substrates. hBN with thickness of ~30 nm was typically chosen. hBN/TBG/hBN stack was assembled and transferred onto 90 nm-thick SiO$_2$ on a silicon substrate, using the previously reported poly-propylene carbonate (PPC)-assisted "tear and stack" dry transfer method[4, 33]. The stack was then annealed at 600 °C for 6 hours. Clean areas free of bubbles and residues were then identified for device fabrication. Device channel was defined using a Vistec 100 kV electron-beam lithography (EBL) system, followed by sulphur hexafluoride ($SF_6$) plasma dry etch. Chromium/gold (4/40 nm) metal contacts were deposited onto the two ends of the channel to form edge contacts[47].

**Low temperature transport measurements.**

The electrical transport measurements down to 83 K were performed in a modified Linkam HFS600E-PB4 cryostat. An Agilent B1500A semiconductor parameter analyser was used to apply gate voltages, source-drain bias, and measure the resistance. We used small $V_{DS}$ of 10 mV during the transport measurements in order to minimize $V_{DS}$-induced doping non-uniformity across the devices. The electrical transport measurements down to 2 K were performed in a Quantum Design physical property measurement system (PPMS) DynaCool. A Stanford Research SR830 lock-in amplifier was used to apply an alternating $I_{DS}$ of 100 nA at 17.777 Hz while simultaneously monitor the in-phase $V_{DS}$. A Keithley 2400 sourcemeter was used to apply the gate voltages.

**Photocurrent measurements and responsivity calculations**

The infrared light (5.0 µm, 7.7 µm, or 12 µm) emitted from a quantum cascade laser was coupled to a Bruker Vertex 70 Fourier transform infrared spectrometer (FTIR) and focused on the sample using a Hyperion 2000 infrared microscope. The sample was placed in the modified Linkam HFS600E-PB4 cryostat, which was mounted on the FTIR stage. The cryostat was filled with argon and was then cooled to 83 K. The light was chopped at 917 Hz by a mechanical chopper. The source-drain bias was applied by a Keithley 2612 sourcemeter, and the photocurrent was collected by means of a Stanford Research SR830 lock-in amplifier in series with a Femto DLPCA-200 current amplifier. The back-gate voltage was applied by the same sourcemeter. A LabVIEW program swept the gate voltage while recording the photocurrent and device resistance.

The responsivity calculations mainly involved the calibrations of incident light power on device. The light power under the infrared microscope was measured by a Thorlabs S401C thermal power sensor. For the 5.0 µm, 7.7 µm, and 12 µm laser light used in the experiments, the powers were 350 µW, 1200 µW, and 1100 µW, respectively. Since our devices (3 µm × 3 µm) are smaller than the beam spot, we need to calibrate the incident power on a device. We calibrated the beam spot size and position using the knife-edge technique, assuming a Gaussian beam shape. For the 5.0 µm laser spot on sample chip, the standard deviations were measured to be $\sigma_x = 6.0$ µm and $\sigma_y = 9.0$ µm along two perpendicular directions, respectively. As a result, by positioning device in the middle of the beam spot, the incident power on the device was calculated to be 9.1 µW. For the 7.7 µm and 12 µm incident light, the same method was applied and the light powers on device were 13 µW and 5.3 µW, respectively. We finally calculated the responsivity using the directly measured photocurrent divided by the power on device.

**Theoretical model for moiré band structures**

We obtain the moiré band structures of TBG with small twist angles by using the 2011 Bistritzer-MacDonald model[3]. In this model, the AA and AB tunnelling amplitudes are $t_{AA} = t_{AB} = $ 110 meV, the Fermi velocity is $v_F = 1 \times 10^6$ m/s, and the flat bands appear at the first magic angle 1.08°. Koshino *et al.* suggested[48] to use $t_{AA} = 79.7$ meV, $t_{AB} = 97.5$ meV, and $v_F = 7.98 \times 10^5$ m/s to accommodate the effect of lattice relaxation. In this work, we use these updated parameters. The momentum cutoff used in these calculations is 6 times as large as the moiré reciprocal lattice vectors. The considered area in the momentum space, which is centred at the $K$ point of the original Brillouin zone (BZ), is 108 times as large as the area of the first moiré BZ. To obtain the optical transitions, we discretize the first moiré BZ into a mesh with 120,000 points (with the hexagonal symmetry intact) and use 1 meV as the integrated energy interval. Each moiré band is spin degenerate, since the spin-orbit coupling is negligibly weak in TBG. The moiré bands at valley $K'$ of the original BZ (not shown) can be readily obtained leveraging the time-reversal symmetry from the bands at valley $K$. As evidenced by many graphene experiments, the two valleys are well decoupled in the bulk yet unavoidably coupled near atomic edges.

**Optical absorption calculations**

During an optical transition event, the momentum is conserved for the initial state $|i>$ and the final state $|f>$, and we ignore the photon momentum. As a result, in our calculations only direct transitions between bands are taken into account, and the calculation processes are similar to those reported in ref. 38. The incident energy flux is $W_i = \frac{c}{4\pi}|\boldsymbol{\Theta}|^2$, and the absorbed energy per unit time is $W_a = \eta\hbar\omega$. Here $\boldsymbol{\Theta}$ and $\omega$ are the electric field and frequency of the light, and the absorption events per unit time $\eta$ can be calculated by using Fermi's Golden rule as $\eta = \left(\frac{2\pi}{\hbar}\right)|M|^2 D$, where $M = <f|H_{int}|i>$ is the matrix element for the light-electron interaction $H_{int}$, and $D$ is the joint density of states at the photon frequency. The light-electron interaction can be obtained by the Hamiltonian $H = v_F \boldsymbol{\sigma} \cdot \left(\boldsymbol{p} - \frac{e}{c}\boldsymbol{A}\right) = H_0 + H_{int}$, which is the same for both graphene layers in the small-twist-angle limit. Here $\boldsymbol{A} = \frac{ic}{\omega}\boldsymbol{\Theta}$ is the vector potential, and $\boldsymbol{\sigma}$ are the Pauli matrices of the two sublattices. $H_0$ is the low-energy Hamiltonian for monolayer graphene, and $H_{int} = v_F \boldsymbol{\sigma} \cdot \frac{e}{i\omega}\boldsymbol{\Theta}$ is the light-electron interaction. The absorption can be expressed by $P = \frac{<W_a>}{W_i}$. The interaction matrix element $<f|\boldsymbol{\sigma} \cdot \boldsymbol{\Theta}|i>$ and the joint density of states $D$ need to be calculated numerically. Note that for monolayer graphene $P = \frac{\pi e^2}{\hbar c}$.

**Figure Captions**

**Fig. 1 | Strong mid-infrared photoresponse in 1.81° twisted bilayer graphene. a**, The schematic of hBN encapsulated TBG device and the photocurrent measurement scheme. **b**, The device source-to-drain resistance as a function of back gate voltage at 83 K. The upper *x*-axis shows the electron filling states. Inset: Calculated bandstructure of the 1.81° TBG. Two superlattice-induced bandgaps above and below the lowest moiré Dirac bands are clearly shown. **c**, The infrared photoresponse of the device under 5.0 µm, 7.7 µm, and 12 µm light illuminations. The extrinsic responsivity is featured by two positive peaks at both hole- and electron-side bandgaps. Other than bandgap positions (and around), the responsivity is negative. For consistency, all curves are horizontally shifted by their respective charge-neutrality point voltages ($V_{CNP}$). In all our devices, the $V_{CNP}$ is within ±2 V. The measurements were performed at 83 K. Inset: The optical image of the 1.81° TBG device. Scale bar: 2 µm.

**Fig. 2 | Bolometric photocurrent enhanced by moiré superlattice. a**, The calculated dynamical conductivity (real part) spectrum of the 1.81° TBG when the Fermi level is tuned to the middle of the superlattice bandgap in the electron branch. Inset: purple, blue, and red arrows schematically indicate some possible interband transitions under 5.0 µm, 7.7 µm, and 12 µm light illuminations, respectively. **b**, Temperature-dependent conductance of the 1.81° TBG. The temperature ranges from 83 K to 300 K. Around $V_{BG}$ of ±43.5 V, the device shows the thermal excited behaviour, indicating the formations of superlattice-induced bandgaps. The $V_{BG}$ scanning range is intentionally reduced at higher temperatures in order to keep a minimal gate leakage current. Inset: The calculated TCC at 83 K, derived from the temperature-dependent conductance. TCC has similar features to the photocurrent patterns.

**Fig. 3 | Photoresponse in bilayer graphene with ultrasmall twist angle. a**, The resistance of the ultrasmall-twist-angle TBG device as a function of back-gate voltage at 2 K. Beside the Dirac point peak, there are two small side ripples denoted by the black arrows. They correspond to electron filling states of ±4, from which the twist angle of 0.37° is obtained. Inset: Calculated bandstructure of the 0.37° TBG. No superlattice bandgap is observed within the plotted energy range. **b**, The extrinsic photoresponsivity of the device under 5.0 µm, 7.7 µm, and 12 µm light illuminations at 83 K. The photoresponse shows a small and positive peak at around the Dirac points, which is distinctively different from that in TBG with bandgap. Inset: TCC calculated from the temperature-dependent transport measurements.

**Competing financial interests**

The authors declare no competing financial interests.

**Acknowledgement**

We acknowledge financial support from the National Science Foundation EFRI-NewLAW program (1741693). We also thank Office of Naval Research for partial support in the experimental setups. The theoretical work at UTD is supported by Army Research Office under Grant No. W911NF-18-1-0416 and Natural Science Foundation under Grant No. DMR-1921581 through the DMREF program. Growth of hexagonal boron nitride crystals was supported by the Elemental Strategy Initiative conducted by the MEXT, Japan and the CREST (JPMJCR15F3), JST. We also acknowledge Dr. Lei Wang, David Hynek, John Woods, Prof. Judy Cha at Yale West Campus and our previous group member Xiaolong Chen for their support.

**Author contributions**

B.D., C.M. and S.Y. fabricated and characterized the devices. Q.W. and F.Z. performed the theoretical calculations. K.W. and T.T. synthesized the hBN crystals. F.X., F.Z., B.D. and Q.W. drafted the manuscript. All the authors discussed the results and commented on the manuscript.

**Data availability**

The data that support the plots within this paper and other findings of this study are available from the corresponding authors upon reasonable request.

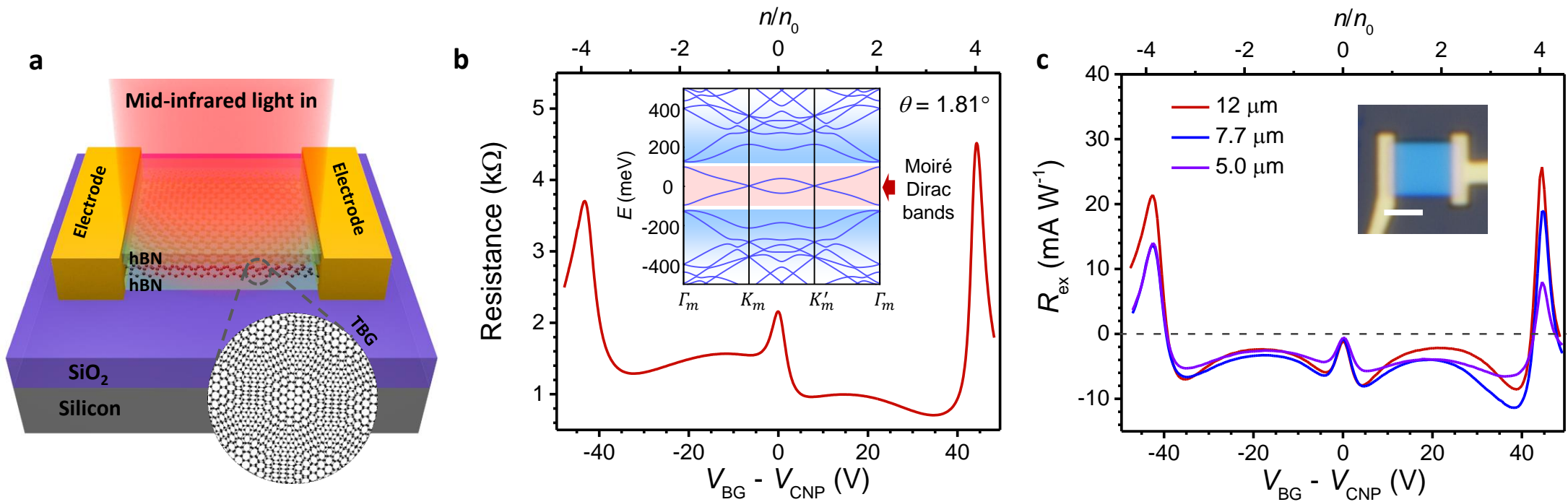

Figure 1: Strong mid-infrared photoresponse in 1.81° twisted bilayer graphene

# Figure 2: Bolometric photocurrent enhanced by moiré superlattice

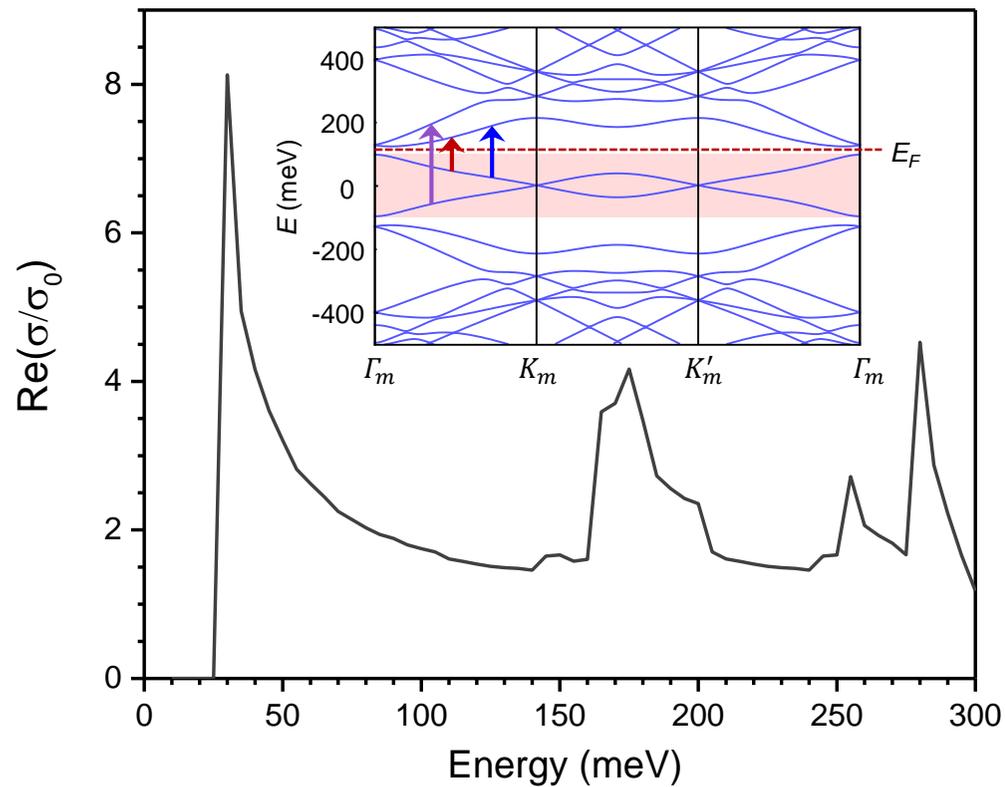
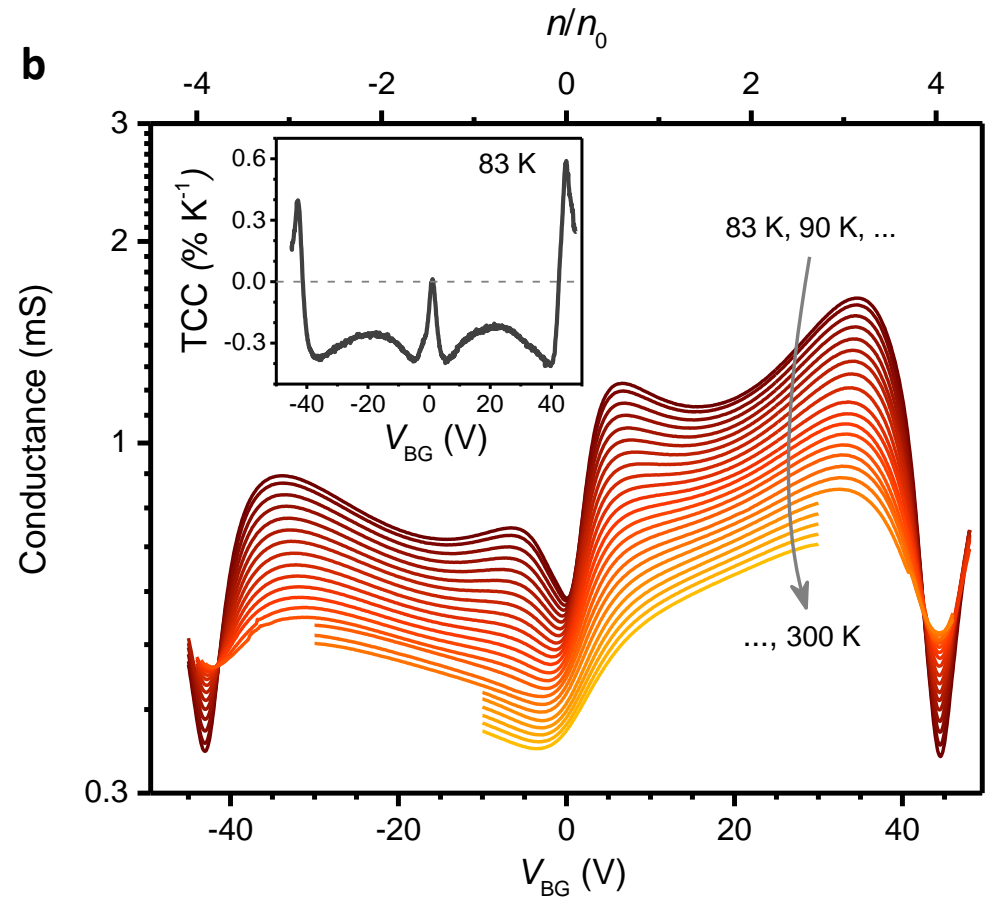

## Figure 3: Photoresponse in bilayer graphene with ultrasmall twist angle

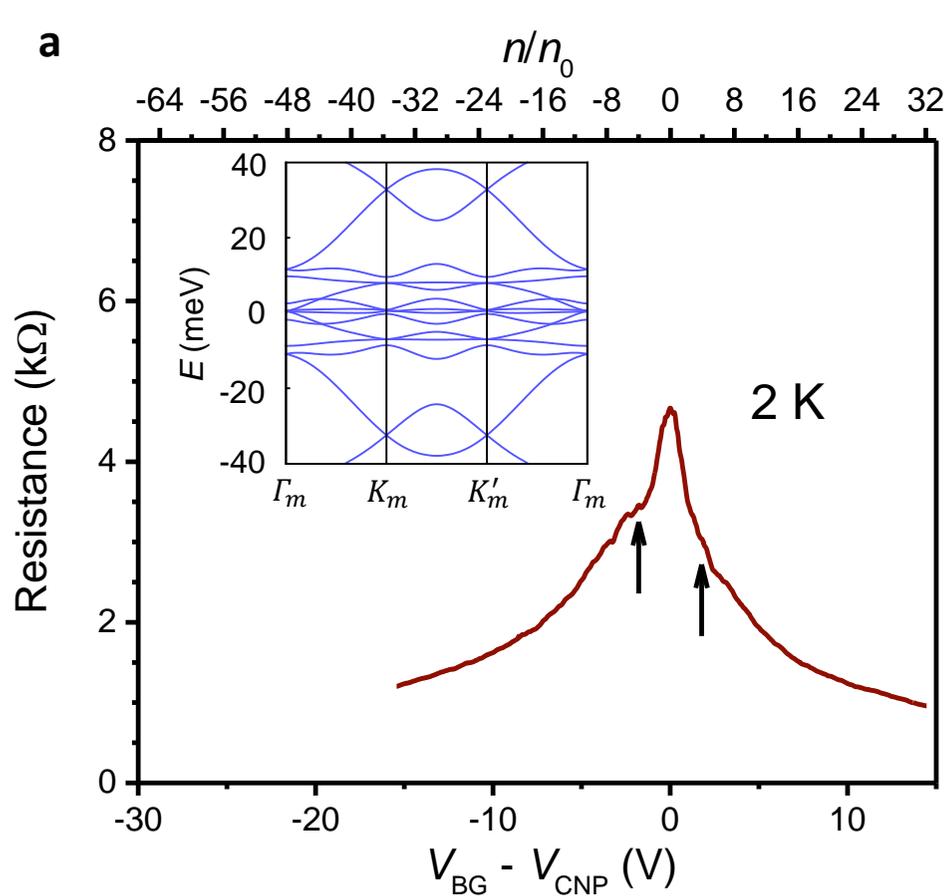
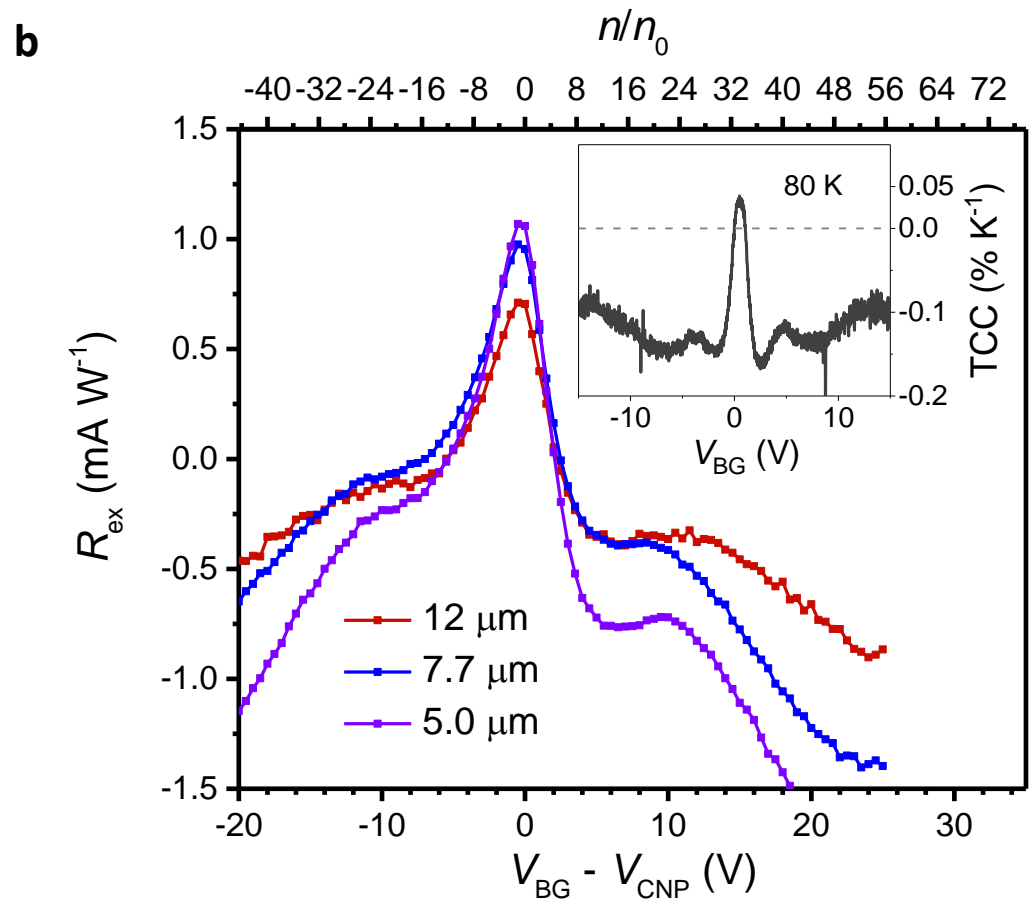